\documentstyle[aps,prd,preprint]{revtex}

\newcommand{\beq}{\begin{equation}}
\newcommand{\eeq}{\end{equation}}
\def\p{\partial}

\def\lap{\lower.5ex\hbox{$\; \buildrel < \over \sim \;$}}
\def\gap{\lower.5ex\hbox{$\; \buildrel > \over \sim \;$}}

\def\L{\Lambda}
\def\rV{\rho_V}
\def\rL{\rho_\L}

\begin{document}

\title{Cosmological constant problems and their
solutions}

\author{Alexander Vilenkin}

\address{
Institute of Cosmology, Department of Physics and Astronomy,\\
Tufts University, Medford, MA 02155, USA}

\maketitle

\begin{abstract}

There are now two cosmological constant problems: (i) why the
vacuum energy is so small and (ii) why it comes to dominate at about
the epoch of galaxy formation.  Anthropic selection appears to be the
only approach that can naturally resolve both problems.  The challenge
presented by this approach is that it requires scalar fields with
extremely flat potentials or four-form fields coupled to branes with
an extremely small charge.  Some recent suggestions are reviewed on
how such features can arise in particle physics models.

\end{abstract}

\section{The problems}

Until recently, there was only one cosmological constant problem
and hardly any solutions.  Now, within the scope of a few years, we have
made progress on both accounts.  We now have two
cosmological constant problems (CCPs) and a number of proposed solutions.
In this talk I am going to review the situation,
focussing mainly on the anthropic approach and on its implications for
particle physics models.
I realize that the anthropic approach has a low approval rating
among physicists.  But I think its bad reputation is largely
undeserved.  When properly used, this approach is quantitative and has
no mystical overtones that are often attributed to it.
Moreover, at present this appears to
be the only approach that can solve both CCPs.  I will also comment on
other approaches to the problems.

The cosmological constant is (up to a factor) the vacuum energy
density, $\rV$.  Particle physics models suggest that the natural
value for this constant is set by the Planck scale $M_P$,
\beq
\rV\sim M_P^4 \sim (10^{18} ~GeV)^4,
\label{Planck}
\eeq
which is some $120$ orders of magnitude greater than the observational
bound,
\beq
\rV\lesssim (10^{-3}~{\rm eV})^4.
\eeq
In supersymmetric theories, one can expect a lower value,
\beq
\rV\sim\eta_{SUSY}^4,
\label{susy}
\eeq
where $\eta_{SUSY}$ is the supersymmetry breaking scale.  However,
with $\eta_{SUSY}\gtrsim 1~$TeV, this is still 60 orders of magnitude
too high.  This discrepancy between the expected and observed values
is the first cosmological constant problem.  I will refer to it
as the old CCP.

Until recently, it was almost universally
believed that something so small could only be zero, due
either to some symmetry or to a dynamical
adjustment mechanism.  (For a review of the early work on CCP, see
\cite{WeinbergPRep}.)
It therefore came as a surprise when recent observations \cite{Supernova}
provided evidence that the universe is accelerating, rather than
decelerating, suggesting a non-zero cosmological constant.  The
observationally suggested value is
\beq
\rV\sim\rho_{M0}\sim (10^{-3}~{\rm eV})^4,
\label{Lobs}
\eeq
where $\rho_{M0}$ is the present density of matter.  This brings yet
another puzzle.  The matter density $\rho_M$ and the vacuum energy
density $\rV$ scale very differently with the expansion of the
universe, and there is only one
epoch in the history of the universe when $\rho_M\sim\rV$.  It is
difficult to understand why we happen to live in this special epoch.
Another, perhaps less anthropocentric statement of the problem is why
the epoch
when the vacuum energy starts dominating the universe ($z_V\sim 1$)
nearly coincides with the epoch of galaxy formation ($z_G\sim 1-3$),
when the giant galaxies were assembled and the bulk of star formation
has occurred:
\beq
t_V\sim t_G.
\eeq
This is the so-called cosmic coincidence problem, or the second CCP.

\section{Proposed solutions}

\subsection{Quintessence}

Much of the recent work on CCP involves the idea of quintessence
\cite{Steinhardt}).  Quintessence
models require a scalar field $\phi$ with a potential $V(\phi)$
aproaching zero at large values of $\phi$.  A popular example is an
inverse power law potential,
\beq
V(\phi)=M^{4+\beta}\phi^{-\beta},
\eeq
with a constant $M\ll M_P$.  It is assumed that initially $\phi\ll M_P$.
Then it can be shown that the quintessence field $\phi$ approaches an
attractor ``tracking'' solution
\beq
\phi(t)\propto t^{2/(2+\beta)}
\eeq
in which its energy density grows relative to that of matter,
\beq
\rho_\phi /\rho_M\sim \phi^2/M_P^2.
\eeq
When $\phi$ becomes comparable to $M_P$, its energy dominates the
universe.  At this point the nature of the solution changes: the
evolution of $\phi$ slows down and the universe enters an epoch of
accelerated expansion.  The mass parameter $M$ can be adjusted so that
this happens at the present epoch.

A nice feature of the quintessence models is that their evolution is
not sensitive to the choice of the initial conditions.  However, I do
not think that these models solve either of the two CCPs.  The
potential $V(\phi)$ is assumed to vanish in the asymptotic range $\phi
\to\infty$.  This assumes that the old CCP has been solved by some
unspecified mechanism.  The coincidence problem also remains
unresolved, because the time of quintessence domination depends on the
choice of the parameter $M$, and there seems to be no reason why this
time should coincide with the epoch of galaxy formation.

\subsection{$k$-essence}

A related class of models involves $k$-essence, a scalar field with a
non-trivial kinetic term in the Lagrangian \cite{k-essence},
\beq
L = \phi^{-2} K[(\nabla\phi)^2].
\eeq
For a class of functions $K(X)$, the energy density of $k$-essence
stays at a constant fraction of the radiation energy density
during the radiation era,
\beq
\rho_\phi/\rho_{rad} \approx {\rm const},
\label{kessence}
\eeq
and starts acting as an effective
cosmological constant with the onset of matter domination.
The function $K(X)$ can be designed so that the constant in
Eq.~(\ref{kessence}) is $\lesssim 10^{-2}$, thus avoiding conflict
with nucleosynthesis, and that $k$-essence comes to dominate at $z\sim
1$.

This is an improvement over quintessence, since
the accelerated expansion in this kind of models always begins during the
matter era.  Galaxy formation can also occur only in the matter era,
but still there seems to be no reason why the two epochs should coincide.
The epoch of $k$-essence
domination $z_V$ is determined by the form of the function $K(X)$, and the
epoch of galaxy formation $z_G$ is determined by the amplitude of primordial
density fluctuations,
\beq
Q=\delta\rho/\rho\sim 10^{-5}.
\eeq
It is not clear why these seemingly unrelated quantities should give
$z_V \sim z_G$ within one order of magnitude.   And of course the old
CCP also remains unresolved.

\subsection{A small cosmological constant from fundamental physics}

One possibility here is that some symmetry of the fundamental physics
requires that the cosmological constant should be zero.
A small value of $\rV$ could then arise due to a small violation of
this symmetry.  One could hope that $\rV$ would be given by an
expression like
\beq
\rV\sim M_W^8/M_P^4\sim (10^{-3}~{\rm eV})^4,
\label{mw}
\eeq
where $M_W\sim 10^3~$ GeV is the electroweak scale.  There have been
attempts in this direction \cite{Arkani}, but
no satisfactory implementation of this program has yet been
developed.  And even if we had one, the time coincidence $t_V\sim t_G$
would still remain a mystery.

Essentially the same remarks apply to the braneworld \cite{braneworld}
and holographic \cite{holography} approaches to CCPs.

\subsection{Anthropic approach}

According to this approach, what we perceive as the
cosmological constant is in fact a stochastic variable which varies on
a very large scale, greater than the present horizon, and takes
different values in different parts of the universe.
We shall see
that situation of this sort can naturally arise in the context
of the inflationary scenario.

The key observation here is that the gravitational clustering that
leads to galaxy formation effectively stops at
$z\sim z_V$.  An anthropic bound on $\rho_V$ can be
obtained by requiring that it does not dominate before the redshift
$z_{max}$ when the earliest galaxies are formed.
With $z_{max}\sim 5$ one obtains \cite{Weinberg87}
\beq
\rho_V\lesssim 200\rho_{M0}.
\label{Wbound}
\eeq
For negative values of $\rV$, a lower bound can be obtained by
requiring that the universe does not recollapse before life had a
chance to develop \cite{Barrow},
\beq
\rho_V\gtrsim -\rho_{M0}.
\eeq

The bound (\ref{Wbound})
is a dramatic improvement over (\ref{Planck}) or (\ref{susy}),
but it still falls short of the observational bound by a factor of about 50.
If all values in the anthropic range (\ref{Wbound}) were equally
probable, then $\rho_V\sim\rho_{M0}$ would still be ruled out at a 95\%
confidence level.
However, the values in this range are {\it not} equally probable.
The anthropic bound (\ref{Wbound}) specifies the value of $\rho_V$
which makes galaxy formation barely possible.
Most of the galaxies will be not in regions characterized by
these marginal values, but rather in regions where $\rV$ dominates
after the bulk of galaxy formation has occured, that is $z_V\lesssim 1$
\cite{AV95,Efstathiou}.

This can be made quantitative by introducing the
probability distribution
as \cite{AV95}
\beq
d{\cal P}(\rV)={\cal P}_*(\rV)\nu(\rV)d\rV.
\label{dP}
\eeq
Here, ${\cal P}_*(\rV)d\rV$ is the prior distribution, which
is proportional to the volume of those parts of the universe where
$\rho_V$ takes values in the interval $d\rV$, and $\nu(\rV)$ is the
average number of galaxies that form per unit volume with a given
value of $\rV$.
The calculation of $\nu(\rV)$ is a standard
astrophysical problem; it can be done, for example, using the
Press-Schechter formalism \cite{PS}.

The distribution (\ref{dP}) gives the probability that a randomly
selected galaxy is located in a region where the effective
cosmological constant is in the interval $d\rho_V$.  If we are typical
observers in a typical galaxy, then we should expect to observe a
value of $\rho_V$ somewhere near the peak of this distribution.

The prior distribution
${\cal P}_*(\rV)$ should be determined from the inflationary model of
the early universe.
Weinberg \cite{Weinberg96,Weinbergcomment}
has argued that a flat distribution,
\beq
{\cal P}_*(\rV)=const,
\label{WC}
\eeq
should generally be a good approximation.  The reason is that the
function ${\cal P}_*(\rV)$ is expected to vary on some large particle
physics scale, while we are only interested in its values in the tiny
anthropically allowed range (\ref{Wbound}).  Analysis shows that this
Weinberg's conjecture is indeed true in a wide class of models, but
one finds that it is not as automatic as one might expect \cite{GV,GV2}.

Martel, Shapiro and Weinberg \cite{MSW} (see also
\cite{Efstathiou,Weinberg96})
presented a detailed calculation of $d{\cal P}(\rV)$ assuming a flat
prior distribution (\ref{WC}).  They found that
the peak of the resulting probability distribution is close to the
observationally suggested values of $\rV$.

The cosmic time coincidence
is easy to understand in this approach \cite{GLV,Bludman}.  Regions of
the universe where $t_V\ll t_G$ do not form any galaxies at
all, whereas regions where $t_V\gg t_G$ are suppressed by ``phase
space'', since they correspond to a very tiny range of $\rho_V$.
It was shown in Ref. \cite{GLV} that the probability
distribution for $t_G/t_V$ is peaked at $t_G/t_V \approx 1.5$, and
thus most observers will find themselves in galaxies formed at
$t_G\sim t_V$.

We thus see that the anthropic approach naturally resolves both CCPs.
All one needs is a particle physics model that would allow $\rV$ to
take different values and an inflationary cosmological
model that would give a more or less flat prior
distribution ${\cal P}_*(\rV)$ in the anthropic range (\ref{Wbound}).

\section{Models with a variable $\rV$}

\subsection{Scalar field with a very flat potential}

One possibility is that what we
perceive as a cosmological constant is in fact a potential $V(\phi)$ of
some field $\phi(x)$ \cite{GV}.  The slope of the potential is assumed to be so
small that the evolution of $\phi$ is slow on the cosmological time scale.
This is achieved if the slow roll conditions
\begin{equation}
 M_P^2 V''\ll V \lesssim \rho_{M0}, \label{sr1}
\end{equation}
\begin{equation}
 M_P V' \ll V \lesssim \rho_{M0}, \label{sr2}
\end{equation}
are satisfied up to the present time.
These conditions ensure that the field is overdamped by the
Hubble expansion, and that the kinetic energy is negligible
compared with the potential energy (so that the equation of state
is basically that of a cosmological constant term.)  The field $\phi$
is also assumed to have negligible couplings to all fields other than
gravity.

Let us now suppose that there was a
period of inflation in the early universe, driven by the potential of
some other field.
The dynamics of the field $\phi$
during inflation are strongly influenced by quantum fluctuations, causing
different regions of the universe to thermalize with different values of
$\phi$.  Spatial variation of $\phi$ is thus a natural outcome of
inflation.

The probability distribution ${\cal P}_*(\phi)$ is determined mainly by the
interplay of two effects.  The first is the ``diffusion'' in the field
space caused by quantum fluctuations.  The dispersion of $\phi$ over a
time interval $\Delta t$ is
\beq
\Delta\phi\sim H(H\Delta t)^{1/2},
\eeq
where $H$ is the inflationary expansion rate.
The effect of diffusion is to make all values of $\phi$ equally
probable over the interval $\Delta\phi$.  The second effect is the
differential expansion.  Although $V(\phi)$ represents only a tiny
addition to the inflaton potential, regions with larger values of $V(\phi)$
expand slightly faster, and thus the probability for higher values of
$V(\phi)$ is enhanced.  The time it takes
the field $\phi$ to fluctuate across the anthropic range
$\Delta\phi_{anth}\sim \rho_{M0}/V'$ is $\Delta t_{anth}\sim
(\Delta\phi_{anth})^2/H^3$, and the characteristic time for
differential expansion is $\Delta t_{de}\sim HM_P^2/V$.

The effect of
differential expansion is negligible if $\Delta t_{anth}\ll \Delta
t_{de}$.  The corresponding condition on $V(\phi)$ is \cite{GV2}
\beq
{V'}^2\gg\rho_{M0}^3/H^3 M_P^2.
\label{C'}
\eeq
In this case, the probability distribution for $\phi$ is flat in the
anthropic range,
\beq
{\cal P}_*(\phi)=const.
\eeq
The probability distribution for the effective
cosmological constant $\rho_V = V(\phi)$ is given by
$$
{\cal P}_*(\rho_V) = {1\over V'} {\cal P_*}(\phi),
$$
and it will also be very flat, since $V'$ is typically almost constant in
the anthropic range. As we discussed in Section II, a flat prior
distribution for the effective cosmological constant in
the anthropic range entails an automatic explanation for the two
cosmological constant puzzles.

On the other hand, if the condition (\ref{C'}) is not satisfied,
then the prior probability
for the field values with a higher $V(\phi)$ would be
exponentially enhanced with respect to the field values at the
lower anthropic end. This would result in a prediction for the
effective cosmological constant which would be too high compared
with observations.

A simple example is given by a potential of the form
\begin{equation}
V(\phi) = \rho_\L + {1\over 2}\mu^2 \phi^2 \label{both},
\end{equation}
where $\rho_\L$ represents the "true" cosmological constant.
We shall assume that $\rho_\L<0$, so that the two terms in
(\ref{both}) partially cancel one another in some parts of the universe.
With $|\rL|\sim (1~{\rm TeV})^4$, the slow roll conditions
(\ref{sr1}), (\ref{sr2}) give
\beq
\mu\lesssim 10^{-90}M_P.
\label{smallmu}
\eeq
Thus, an exceedingly small mass scale must be introduced.

The condition (\ref{C'}) yields a lower bound on $\mu$,
\beq
\mu\gtrsim 10^{-137} M_P.
\eeq
Here, I have used the upper bound on the expansion rate at late stages
of inflation, $H\lesssim 10^{-5}M_P$, which follows from the CMB
observations.

We thus see that models with a variable $\rho_V$ can be
easily constructed in the framework of inflationary cosmology.
The challenge here is to explain the very small mass scale
(\ref{smallmu}) in a natural way.

\subsection{Four-form models}

Another class of models, first discussed by Brown and
Teitelboim \cite{Teitelboim}, assumes that the cosmological constant is due to
a four-form field \cite{Duff},
\beq
F^{\alpha\beta\gamma\delta}=F\epsilon^{\alpha\beta\gamma\delta}.
\label{F}
\eeq
The field equation for $F$ is $\p_\mu F=0$, so $F$ is a constant, but it
can change its value through nucleation of bubbles bounded by domain
walls, or branes.  The total
vacuum energy density is given by
\beq
\rV=\rL +F^2/2
\label{rhoF}
\eeq
and once again it is assumed that $\rL<0$.
The change of the field across the brane is
\beq
\Delta F=q,
\label{Fq}
\eeq
where the ``charge'' $q$ is a constant fixed by the model.  Thus, $F$
takes a discrete set of values, and the resulting spectrum of $\rV$ is
also discrete.  The four-form
model has recently attracted much attention
\cite{Bousso,Donoghue,FMSW,Banks,GV2} because four-form fields
coupled to branes naturally arise in the context of
string theory.

In the range where the bare cosmological constant is almost
neutralized, $|F|\approx |2\rL|^{1/2}$, the spectrum of $\rV$ is
nearly equidistant, with a separation
\beq
\Delta\rV\approx |2\rL|^{1/2}q.
\label{Deltarho}
\eeq
In order for the anthropic explanation to work, $\Delta\rV$ should not exceed
the present matter density,
\beq
\Delta\rV\lesssim\rho_{m0}\sim (10^{-3}~eV)^4.
\label{ebound}
\eeq
With $\rL\gtrsim (1~TeV)^4$, it follows that
\beq
q\lesssim 10^{-90}M_P^2.
\label{qbound}
\eeq
Once again, the challenge is to find a natural explanation for such
very small values of $q$.


In order to solve the cosmological constant problems, we have to
require in addition that {\it (i)} the probability distribution for
$\rV$ at the end of inflation is nearly flat, ${\cal P}_*(\rV)\approx
const$, and {\it (ii)} the brane nucleation rate is sufficiently low,
so that the present vacuum energy does not drop significantly in less
than a Hubble time.  Models satisfying all the requirements can be
construcled, but the conditions {\it (i), (ii)} significantly
constrain the model parameters.  For a detailed discussion, see
\cite{GV2}.

\section{Explaining the small parameters}

Both scalar field and four-form models discussed above have some
seemingly unnatural features.  The scalar field models require
extremely flat potentials and the four-form models require branes with
an exceedingly small charge.  The models cannot be regarded as
satisfactory until the smallness of these parameters is explained in a
natural way.  Here I shall briefly review some possibilities that
have been suggested in the literature.

\subsection{Scalar field renormalization}

Let us start with the scalar field model.
Weinberg \cite{Weinbergcomment}
suggested that the flatness of the potential could
be due to a large field renormalization.
Consider the Lagrangian of the form
\beq
L={Z\over{2}}(\nabla\phi)^2 - V(\phi).
\eeq
The potential for the canonically normalized field $\phi'
=\sqrt{Z}\phi$ will be very flat if the field renormalization constant
is very large, $Z\gg 1$.

More generally, the effective Lagrangian for $\phi$ will include
non-minimal kinetic terms \cite{Donoghue,GV2},
\beq
L={1\over{2}}F^2(\phi)(\nabla\phi)^2-V(\phi).
\label{exp}
\eeq
Take for example $F=e^{\phi/M}$.  Then the potential for the canonical
field $\psi=Me^{\phi/M}$ is $V(M\ln(\psi/M))$.  This will typically
have a very small slope if $V(\phi)$ is a polynomial function.
It would be good to have some particle physics motivation either for a
large running of the field renormalization, or for an exponential
function $F(\phi)$ in the Lagrangian (\ref{exp}).

\subsection{A discrete symmetry}

Another approach attributes the flatness of the potential to a
spontaneously broken discrete symmetry \cite{Gia}.
The main ingredients of the model are: (1) a
four-form field $F_{\mu\nu\sigma\tau}$ which can be obtained from a
three-form potential,
$F_{\mu\nu\sigma\tau}=\p_{[\mu}A_{\nu\sigma\tau]}$, (2) a complex
field $X$ which develops a vacuum expectation value
\beq
\langle X\rangle=\eta e^{ia}
\eeq
and whose phase $a$ becomes a Goldstone boson, and (3) a scalar field
$\Phi$ which is used to break a discrete $Z_{2N}$ symmetry.

The action is assumed to be invariant under the following three
symmetries: (1) $Z_{2N}$ symmetry under which
\beq
\Phi\rightarrow\Phi e^{i\pi/N}, ~~~~~~ a\rightarrow -a ~~~ ({\rm or}~~
X\rightarrow X^\dagger),
\label{Z2N}
\eeq
(2) a symmetry of global phase transformations
\beq
a\rightarrow a + {\rm const},
\eeq
and (3) the three-form gauge transformation
\beq
A_{\mu\nu\sigma}\rightarrow A_{\mu\nu\sigma} +\p_{[\mu}B_{\nu\sigma]},
\eeq
where $B_{\nu\sigma}$ is a two-form.  Below the symmetry breaking
scales of $X$ and $\Phi$,
the effective Lagrangian for the model can be written as
\beq
L=\eta^2(\p_\mu a)^2-{1\over{4}}F^2 +({\rm effective~ interactions}).
\label{LaA}
\eeq
The interactions generally include all possible terms that are
compatible with the symmetries.  Among such terms is
the mixing of the Goldstone $a$ with the three-form potential,
\beq
g\eta^2{\langle\Phi\rangle^N\over{M_P^N}}\epsilon^{\mu\nu\sigma\tau}
A_{\nu\sigma\tau} \p_\mu a,
\label{mixing}
\eeq
where $g\lesssim 1$ is a dimensionless coupling and
I have assumed that the Planck scale $M_P$ plays the role of the
ultraviolet cutoff of the theory.

The effect of the mixing term (\ref{mixing}) is to give a mass
\beq
\mu=g\eta{\langle\Phi\rangle^N\over{M_P^N}}
\eeq
to the field $a$.  This mass can be made very small if
$\langle\Phi\rangle \ll M_P$ and $N$ is sufficiently large.  For
example, with $\langle\Phi\rangle\sim 1~$TeV, $\eta\ll M_P$ and $N\geq
6$, we have $\mu\ll 10^{-90}M_P$, as required.

Models of this type can also be used to generate branes with a very
small charge.  In this case $a$ is assumed to be a pseudo-Goldstone
boson, like the axion, and the theory has domain wall solutions
with $a$ changing by $2\pi$ across the wall.  The mixing of $a$ and
$A$ couples these walls to the four-form field, and it can be shown
that the corresponding charge is
\beq
q=2\pi g\eta^2{\langle\Phi\rangle^N\over{M_P^N}}.
\eeq
Once again, the anthropic constraint on $q$ is satisfied for
$\langle\Phi\rangle\sim 1~$TeV, $\eta\ll M_P$ and $N\geq
6$.

The central feature of this approach is the $Z_{2N}$ symmetry
(\ref{Z2N}).  What makes this symmetry unusual is that the phase
transformation of $\Phi$ is accompanied by a charge conjugation of
$X$.  It can be shown, however, that such a symmetry can be naturally
embedded into a left-right symmetric extension of the standard model
\cite{Gia}.

\subsection{String theory inspired ideas}

Feng {\it et. al.} \cite{FMSW}
have argued that branes with extremely small charge
and tension can naturally arise due to non-perturbative effects in
string theory.  A potential problem with this approach is that the
small brane tension and charge appear to be unprotected against
quantum corrections below the supersymmetry breaking scale
\cite{Gia}.  The cosmology of this model is also problematic, since it
is hard to stabilize the present vacuum against copious brane
nucleation \cite{GV2}.

A completely different approach was taken by Bousso and Polchinski
\cite{Bousso}.  They assume that several four-form fields $F_i$ are
present so that the vacuum energy is given by
\beq
\rV=\rho_\L+{1\over{2}}\sum_i F_i^2.
\eeq
The corresponding charges $q_i$ are not assumed to be very
small, but Bousso and Polchinski have shown that with multiple
four-forms the
spectrum of the allowed values of $\rV$ can be sufficiently dense
to satisfy the anthropic condition (\ref{ebound}) in the range of
interest. However, the situation here is quite different from
that in the single-field models.
The vacua with nearby values of $\rV$ have very
different values of $F_i$, and
there is no reason to expect the prior
probabilities for these vacua to be similar. Moreover,
the low energy physics in different vacua is likely to be
different, so the process of galaxy formation and the types of
life that can evolve will also differ.  It appears therefore that
the anthropic approach to solving the cosmological constant
problems cannot be applied to this case \cite{Banks}.

\section{Concluding remarks}

In conclusion, it appears that the only approach that can solve both
   cosmological
   constant problems is the one that attributes them to anthropic
   selection effects.  In this approach what we perceive as the
   cosmological constant is in fact a stochastic variable which
   varies from one part of the univers to another.  A typical observer
   then finds himself in a region with a small cosmological constant
   which comes to dominate at about the epoch of galaxy formation.
Cosmological models of this sort can easily be constructed in the framework of
   inflation.  What one needs is either a scalar
   field with a very flat potential, or a four-form field coupled to
   branes with a very small charge.  Some interesting suggestions have
   been made on how such features can arise; the challenge here is to
   implement these suggestions in well motivated particle physics
   models.  (One attempt in this direction has been made in \cite{Gia}.)

There are also problems to be addressed on the astrophysical side of
the anthropic approach.  All anthropic calculations of the probability
distribution (\ref{dP}) for $\rV$ assumed that observers are in giant
galaxies like ours and identified $\nu(\rV)$ in Eq. (\ref{dP}) with the
density of such galaxies.  This, however, needs some
justification\footnote{I am grateful to David Spergel for emphasizing
this to me.}.  In the hierarchical structure formation scenario, dwarf
galaxies could form as early as $z=10$, and if they are
included among the possible sites for observers, then the expected
epoch of vacuum domination would be $z_V\sim 10$.

One problem with dwarf galaxies is that if the mass of a galaxy is
too small, then it cannot retain the heavy elements dispersed in
supernova explosions.  Numerical simulations suggest that the fraction
of heavy elements retained is $\sim 30\%$ for
a $10^9 M_\odot$ galaxy and is negligible for much smaller galaxies
\cite{LF}.  The heavy elements are necessary for the formation of
planets and of observers, and thus one has to require that the
structure formation hierarchy should evolve up to mass scales $\sim
10^{9} M_\odot$ or higher prior to vacuum domination.

Another point to note is that smaller galaxies formed at earlier times
have a higher density of matter.  If this translates into a higher
density of stars (or dark matter clumps),
then we may have additional constraints by requiring
that the timescales for disruption of planetary orbits by stellar
encounters should not be too short.  However, the
cross-section for planetary orbit disruption is rather
small (comparable to the size of the Solar system), and since close
stellar encounters are quite rare in our galaxy, one does not expect a
large effect from a modest density enhancement in dwarf galaxies.

An interesting possibility is that disruption of orbits of comets,
rather than planets, could be the controlling factor of anthropic
selection \cite{Spergel}.  Comets move around the Sun, forming the
Oort cloud of radius $\sim 0.1$ pc (much greater than the Solar
system!).  Whenever a star or a molecular cloud passes by, the
orbits of some comets are disrupted and some of them enter the
interior of the Solar system.  Occasionally they hit planets, causing
mass extinctions.  The time it took to evolve intelligent beings after
the last major hit is comparable to the typical time interval between
hits on Earth ($\sim 10^8$ yrs), so one could argue that a substantial
increase in the rate of hits might interfere with the evolution of
observers.  There are, of course, quite a few blanks to be filled in
this scenario, and at present we are far from being able to reliably
quantify the scale of bound systems to be used in the definition of
$\nu(\rV)$.  However, if the anthropic approach is on the right track,
then one can {\it predict} that future research will show the relevant
scale to be that of giant galaxies.

Finally, I would like to mention the possibility
of a `compromise' solution to CCPs.  It is conceivable that the cosmological
constant will eventually be determined from the fundamental theory.
For example, it could be given by the relation (\ref{mw}).
This would solve
the old CCP.  The time coincidence problem could then be solved
anthropically if the amplitude of density fluctuations $Q$ is a
stochastic variable.  With some mild assumptions
about the probability distribution ${\cal P}_*(Q)$, one finds that
most galaxies will form at about the time of vacuum domination
\cite{GV2}.

\section{Acknowlegements}

I am grateful to Mario Livio and Paul Frampton for inviting me to the
very interesting meetings where this work was presented and to Gia
Dvali, David
Spergel and Steve Barr for useful dicsussions.  This work was
supported in part by the National Science Foundation and by the
Templeton Foundation.

\end{document}